# Exploring Spatial Patterns of Interurban Passenger Flows Using Dual Gravity Models

Zihan Wang, Yanguang Chen

(Department of Geography, College of Urban and Environmental Sciences, Peking University, Beijing 100871, P.R. China. E-mail: zihanw@pku.edu.cn; chenyg@pku.edu.cn)

**Abstract:** Passenger flows in a traffic network reflect spatial interaction patterns in an urban systems. Gravity models can be employed to quantitatively describe and predict spatial flows. However, how to model passenger flows and reveal the deep structure of urban and traffic networks in the case of missing partial data is still a problem to be solved. This paper is devoted to characterizing the interurban passenger flows in the Beijing-Tianjin-Hebei region of China by means of dual gravity models and Tencent location big data. The method of parameter estimation is the least squares regression. The main results are as follows. First, both railway and highway passenger flows can be effectively described by the dual gravity model. A small part of missing spatial data can be made up by the predicted values. Second, the fractal properties of traffic flows can be revealed. The railway passenger flows more follow gravity scaling law than the highway passenger flows. Third, the prediction residuals indicate the changing trend of interurban connections in the study area in recent years. The center of gravity of spatial dynamics seems to shift from the Beijing-Tianjin-Tangshan triangle to the Beijing-Baoding-Shijiazhuang axis. A conclusion can be reached that the dual gravity models is an effective tools of for analyzing spatial structures and dynamics of traffic networks and flows. Moreover, the models provide a new approach to estimate fractal dimension of traffic network and spatial flow patterns.

# 1 Introduction

In the research of urban and regional development, the prediction and analysis of spatial flow are very important. According to the idea from a new science of cities advocated by Batty (2013), cities should be treated not only as places in space but also as systems of networks and flows. To understand space, we must understand flows, and to understand flows, we must understand networks as well as the relations between objects that comprise the urban system (Batty, 2013). The new science of cities involves fractal geometry, allometric analysis, and network science (Batty, 2008). The analysis of flows involves not only urban geography, but also transportation geography. A regional network of transportation represents an intersection of urban and traffic sciences. Spatial flow quantity is an important measurement of traffic and urban networks. If a spatial dataset for flows is complete for an urban system, we can adopt spatial interaction models based on entropy maximization to predict interurban flow quantity. However, this model, developed by Wilson (1970), is based on an aim for future spatial optimization rather than an aim for understanding present reality. In particular, Wilson's models are not available in the case of key data missing. In this instance, dual gravity models can be employed to describe and predict spatial flows between cities. This pair of models can be derived from Wilson's spatial interaction models and can be associated with fractal concepts (Chen, 2015). Even if there is data missing, the models can be utilized to describe and predict intra-urban and interurban spatial flows.

Data and models are both necessary for our understanding of urban and traffic networks. The success of science development rests in great emphasis on the role of quantifiable data and their interplay with models in scientific research (Louf and Barthelemy, 2014). Nowadays, big data of passenger flows can reflect the spatial interaction intensity in an urban system correctly and timely (Yan et al, 2009). Spatial interactions in an urban system produce various spatial flows including passenger flows, material flows, information flows, and fund flows, then map them in cyberspace (Jin et al., 2015). Spatial pattern characteristics of passenger flow big data are closely related to the development history, regional status, and future planning of an urban system. This paper is devoted to exploring spatial patterns of interurban passenger flows based on big data from the ~~prospective~~ perspective of traffic networks. The method is based on urban gravity modeling and ideas from fractals. The Beijing-Tianjin-Hebei region in China is taken as a study area in this work due to its typicality and particularity. The railway and highway passenger big data used in this article came from Tencent location big data. The dual gravity models are employed to solve the following problems: characterizing the spatial flows of railway and highway passengers in our study area, estimating the fractal dimension of traffic networks based on flows, predicting the flow quantities of passengers and making up the missing data, analysis the structural and dynamic features of spatial flows. By fitting the dual gravity models with Tencent location big data, we modeled the spatial pattern of passenger flows. Using the residuals and outliers predicted by the models, we compared the gaps between the expectation and reality and researched the dynamical mechanism. The aim is to develop the methodology of geo-spatial analysis of spatial flows in the case of incomplete datasets. The rest parts of this paper are organized as follows: In Section 2, the model, methods, study area, and foundational data are illuminated. In Section 3, the results and analytic process are presented. In Sections 4 and 5, several questions are discussed, and the discussion is concluded by summarizing the main points of this study.

## 2 Methods and data

### 2.1 Dual gravity models

The dual gravity models will be fitted to the observational data to reveal spatial characteristics and dynamical mechanisms of railway and highway passenger flows in the Beijing-Tianjin-Hebei Region. Gravity models are basic form of spatial interaction modeling. Spatial interaction models play a key role in urban geography, and they are widely used in passenger/cargo flow planning, regional trade potential analysis, and urban systems research (Goh *et al.*, 2012; Gu and Pang, 2008; Martinez-Zarzoso, 2003; Tong *et al.*, 2018). The common forms include Reilly's law of retail gravitation (Reilly, 1929), the geographical gravity model by analogy with Newton's law of universal gravity, S.A. Stouffer's intervening-opportunity model (Stouffer, 1940), J.Q. Stewart's potential model (Stewart, 1948), Converse's breaking-point model (Converse, 1949), and Wilson's entropy maximizing model (Wilson, 1969). Theoretically, those models can be divided into 2 categories: one is the gravity models, the other is spatial interaction models. The law of retail gravitation, the potential model, and the breaking-point model have been proved to be deduced from the geographical gravity model (Chen, 2015). Based on the entropy-maximizing hypothesis, Wilson built and optimized a series of spatial interaction models (Wilson, 1968; Wilson, 1969; Wilson, 1970; Wilson, 2000; Wilson, 2010). The gravity models were born earlier but encountered many problems during development. For example, it was not objective to determine the gravitational constant and the distance exponent would change over time (Mikkonen and Luoma, 1999). In the early years, the distance parameter observed in gravity models was a random value between 0 and 3. That phenomenon did not conform to Euclidean geometry and made many geographers confused. The distance-decay function in Wilson's spatial interaction models is exponential law and there is no fractal dimension dilemma. Meanwhile, the parameters in Wilson's models were not obtained by analogy but deduced with some rigorous process. After being published, Wilson's models were highly praised and widely used in geography (Gould, 1972). Naturally, exponential law would be considered a better choice to take place of power law in the distance-decay function. However, there was a characteristic scale $r_0$ in exponential law, which highly relies on research extent. Besides, exponential law implies *locality* rather than *action at a distance*, and long-distance action is the necessary prerequisite for the first law of geography (Chen, 2015). So, the transmission range of gravity in exponential law is very limited. That conflicts with the first law in Geography. Now the dimension dilemma in power law can be solved by fractal theory. The distance exponent in power law has been proved to be a scaling exponent equivalent to the average fractal dimension of cities' size measurements within a region (Chen and Huang, 2018a). However, the locality problem in exponential law hasn't been solved yet.

The traditional geographical gravity model, which is local, describes the relationship between one point and other points. For example, Mackay (1958) used the gravity model to fit phone data for describing the relationship between Montreal and other surrounding cities. Chen (2015) proposed the improved global gravity model with a symmetrical expression, which is termed dual gravity model for global spatial interaction analysis. That model can be used to describe the relationship between any two cities in a geographical region. The equations are as follows

$$T_{ij} = K P_i^u P_j^v r_{ij}^{-\sigma}, \qquad (1)$$

$$T_{ji} = K P_i^v P_j^u r_{ji}^{-\sigma}, \qquad (2)$$

where $T_{ij}$ denotes the intensity of the flow from city $i$ to city $j$, $T_{ji}$ denotes the intensity of the flow from city $j$ to city $i$, $P_i$ denotes the size of city $i$, $P_j$ denotes the size of city $j$, $r_{ij}$ and $r_{ji}$ denote the distance between city $i$ and city $j$. As for the parameters, $K$ is a constant, $u$, $v$, and $\sigma$ are actually gravity exponents, representing a set of cross scaling exponents. From equations (1) and (2) it follows

$$I_{ij} = (T_{ij}T_{ji})^{1/(u+v)} = K^{2/(u+v)} \frac{P_i P_j}{r_{ij}^{2\sigma/(u+v)}} = G \frac{P_i P_j}{r_{ij}^b}, \tag{3}$$

in which $I_{ij}$ denotes the attractive force. The expressions of the parameters in equation (3) follow

$$G = K^{2/(u+v)}, \tag{4}$$

$$b = \frac{2\sigma}{u+v}, \tag{5}$$

where $G$ denotes the gravity coefficient, and $b$ denotes the distance decay exponent, which proved to be the average fractal dimension of the corresponding urban size measurement (Chen, 2015; Chen and Huang, 2018a). In this sense, equations (1), (2), and (3) are in fact fractal gravity models. The flows can reflect the size of the attractive forces between cities, but flows and attractive forces are not equal. The most intuitive performance of the differences between them is: the flows are asymmetrical, while the attractive forces are symmetrical. Specifically, $T_{ij}$ is always not equal to $T_{ji}$, while $I_{ij}$ is equal to $I_{ji}$. The asymmetry of the flows is embodied in the dual gravity model as the difference between the scaling exponents, $u$ and $v$. The essence is that the sizes of the origination city and the destination city have different influence weights on the size of flows. The improvement of the dual gravity model lies in the clear distinguishment of flows from attractive forces in that way.

The theoretical foundation of the dual gravity models lies in the principle of entropy maximization. Based on the allometric scaling relation between city size and spatial flow quantity, equations (1) and (2) can be derived from the revised spatial interaction model of Wilson (1968, 1970). Wilson's spatial interaction models can be derived from the postulate of entropy maximization (Wilson, 1970; Wilson, 2010). This suggests that allometric scaling and entropy maximizing processes are basic rationales of dual gravity models. In fact, entropy maximization ~~are~~ is a very important principle in geographical analysis (Batty, 2010; Batty and Longley, 1994; Batty et al, 2014; Bussiere and Snickers, 1970; Chen, 2012; Chen and Huang, 2018b; Curry, 1964).

The model parameter values can be estimated by least squares calculation. Converting Eq. (1) into logarithmic form, we have

$$\ln T_{ij} = \ln K + u \ln P_i + v \ln P_j - \sigma \ln r_{ij}. \tag{6}$$

In this way can we use $\ln T_{ij}$ as the dependent variable, $\ln P_i$, $\ln P_j$, and $\ln r_{ij}$ as independent variables for multiple linear regression. After getting the model's parameters by least squares calculation, we re-forecasted the passenger flow matrix. Some information about the spatial pattern of regional migrations in the Beijing-Tianjin-Hebei Region was hidden in the residuals between the expected values and the real values.

## 2.2 Study area and data

The study area includes Beijing Municipality, Tianjin Municipality, and Hebei province. The network of cities in the Beijing-Tianjin-Hebei region composed one of the three major urban systems in mainland China (Figure 1). They have considerable influences on the economic development of north China and even the whole country. The unique political advantages and national radiation capabilities of the capital Beijing have made the hierarchy system and the spatial network of the region very different from the other two urban systems (the Yangtze River Delta region and the Pearl River Delta region) (Lu, 2015). At present, the coordinated development of the Beijing-Tianjin-Hebei region has become a Chinese national strategic plan, which will provide reference paradigms for the development of other urban systems. In 2016, *the Beijing-Tianjin-Hebei National Economic and Social Development Plan during the 13th Five-Year Plan period* was issued. The plan put forward 9 tasks for improving the overall development of that urban system. One of the 9 tasks was to focus on accelerating the construction of integrated transportation construction, and 'build the Beijing-Tianjin-Hebei on the track'. The ~~railway~~ transportation network in the Beijing-Tianjin-Hebei region simultaneously assumes the dual functions of intercity transportation within the region and interprovincial transportation outside the region. It is also the core of the national transportation network (Jin and Wang, 2004).

The main data used in this article came from the Tencent location big data platform (https://heat.qq.com/). Tencent location big data, which can reflect destinations and scales of population flows between cities, is generated in real-time by the built-in location calls of Tencent applications such as WeChat, QQ, and Tencent Maps. Location software development kits (SDKs) are embedded in Tencent applications. A location SDK is a set of location-based services (LBS) positioning interfaces that can use positioning application programming interface (API) to obtain positioning results, inverse geocoding, and geofencing. When users of those applications actively call the location SDKs, data will be generated on Tencent servers. The data were processed by specific modules such as distributed Data Warehouse, Real-time Computing, and Data Bank. Then, real-time big data with accurate location can be obtained. They can be used for personal travel planning, business logistics planning, municipal transportation planning, and disaster relief planning. At present, Tencent applications have about 800 million daily active users which covers more than 70% Chinese total population. Location calls per day exceed 55 billion times. Location check-in times per day exceed 60 million. Among similar big data, Tencent location big data is more accurate and reliable (Li *et al.*, 2016). In literature, Tencent location big data has been used to study the real-time population flow networks between cities and provinces. The research methods include time series analysis, complex network analysis, rank-size distribution, and connection intensity (Hu, 2019; Liang *et al*, 2019; Zhou and Xiao, 2018). Besides, Tencent location big data before and after the Spring Festival was used to study the phenomenon of large-scale population migrations in China (Li *et al*., 2016). That data was also used to characterize population distribution and mixed-use buildings at fine scales (Liu *et al*., 2018; Yao *et al*., 2017).

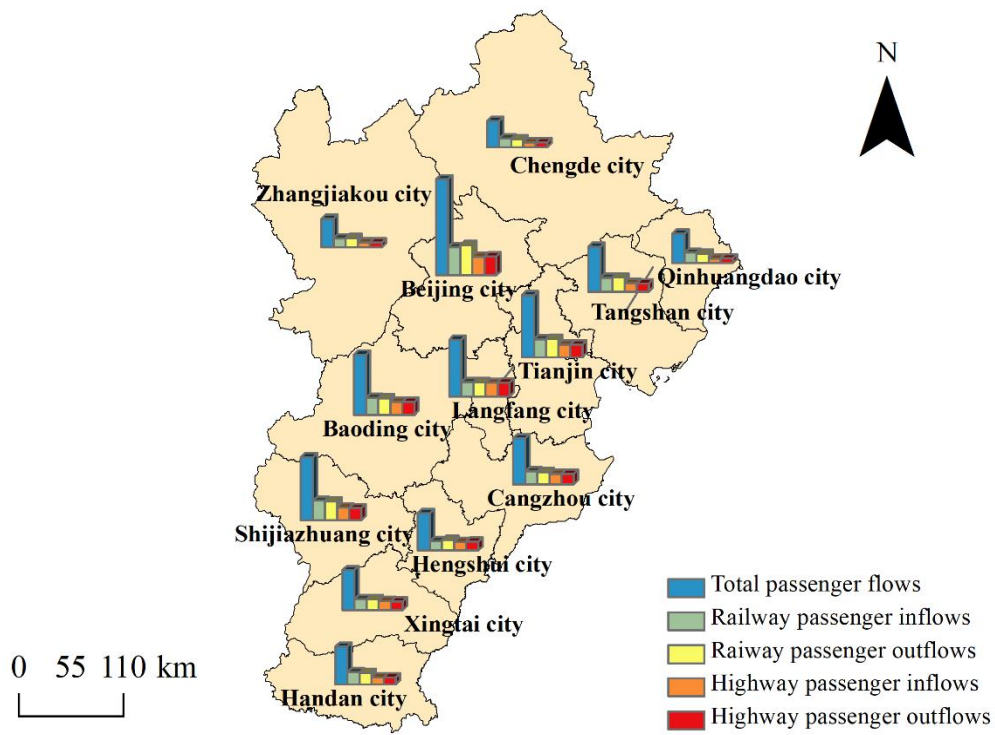

**Figure 1 The spatial pattern of passenger flows in the Beijing-Tianjin-Hebei region**
**Note**: Passenger flows are the sum of railway passenger flows and highway passenger flows.

The passenger flows provided by Tencent location big data describe the intensity of daily intercity population migrations. The passenger flows are calculated comprehensively by passengers' numbers, transportation modes, and travel distance. The flows can be divided into 3 categories according to transportation modes: air passenger flows, railway passenger flows, and highway passenger flows. We didn't collect the air passenger flows, because the travel distance in the Beijing-Tianjin-Hebei region is limited. The actual air passenger flows' values are always zero. The weight of travel distance can't be excluded from flows, but the impact of the weight on analytical results is not significant. Because the travel distance in the region is very slightly compared to nationwide. The population data used in this article is the urban annual average population provided by *China City Statistical Year Book 2017* (See File 1 for processed datasets).

## 2.3 Data processing

After selecting transportation modes, we collected data on the 1st of each month from April 2018 to March 2019 and recorded it in several matrices. To avoid random disturbance, we averaged the corresponding rows and columns in the matrices to get an average highway passenger flow matrix and an average railway passenger flow matrix. During data collecting, it was found that the data set contained obvious outliers. The outliers are passenger flows during the Spring Festival, May Day, and Chinese National Day holidays because large number of Chinese choose to travel during these holidays. These peak passenger flows are 50% more than the normal flows. Data cleaning is therefore necessary for that dataset. However, the variation of the goodness of fit and the regression parameters after data cleaning can be ignored (Table 1). That is to say, the number of collected samples is large enough to make some outliers not affect the final fitting effect of models to data. Besides, data cleaning reduced the number of abnormal predicted values exceeding 2 times the standard deviation range. So data cleaning was included in the follow-up study.

Table 1. The goodness of fit and the regression parameters before and after data cleaning

| Data cleaning | Multiple $R$ | Regression coefficient | | |
|---|---|---|---|---|
| | | $\ln P_i$ | $\ln P_j$ | $\ln r_{ij}$ |
| **After cleaning** | 0.8620 | 0.2374 | 0.2176 | -0.4412 |
| **Before cleaning** | 0.8622 | 0.2356 | 0.2161 | -0.4335 |

**Data Resources**: Railway passenger flows provided by Tencent location big data.

## 3 Results

### 3.1 The spatial pattern of passenger flows in the study area

In light of the average passenger flow matrix, the spatial pattern of the regional passenger flows could be drawn. The results are displayed in Figure 1. As can be seen, the passenger flows of Beijing city, Tianjin city, and Shijiazhuang city which is the capital of Hebei province are at the forefront of the Beijing-Tianjin-Hebei region, and these three cities are the cores of regional population migrations. The passenger flows of Qinhuangdao city, Zhangjiakou city, and Chengde city are the least in the region. These three cities are not only located at the edge of the region but also the trough of regional population migrations. From an overall point of view, the northern and southern ends of the region are passenger flow valleys, and the middle section has more passenger flows. Beijing and its main external connection destinations which are southeast and southwest are passenger flow peaks (Table 2).

**Table 2. Partial processed results for railway passenger flows of Beijing-Tianjin-Hebei region (average values)**

| Passenger flows | Beijing | Tianjing | Shijiazhuang | Tangshan | Qinhuangdao | Handan | Xingtai | Baoding | Zhangjiakou | Chengde | Cangzhou | Langfang | Hengshui |
|---|---|---|---|---|---|---|---|---|---|---|---|---|---|
| Beijing | 0 | 19.0700 | 14.3900 | 10.5200 | 9.7900 | 11.2500 | 9.4300 | 18.5000 | 13.2800 | 10.0800 | 9.3200 | 21.1100 | 8.1222 |
| Tianjing | 18.1900 | 0 | 7.3100 | 10.5700 | 6.7800 | 5.3500 | 4.7000 | 7.5200 | 4.3000 | 3.8600 | 9.4100 | 9.7900 | 4.8222 |
| Shijiazhuang | 13.0200 | 7.1667 | 0 | 5.9700 | 4.8444 | 9.7500 | 11.3300 | 13.0800 | 4.5300 | 3.7300 | 6.2600 | 5.9000 | 7.3600 |
| Tangshan | 9.8900 | 10.4400 | 6.2400 | 0 | 8.6200 | 5.6000 | 4.3500 | 4.9300 | 3.0800 | 4.7500 | 4.2000 | 5.0500 | 2.9333 |
| Qinhuangdao | 9.0500 | 6.6600 | 5.0500 | 8.5800 | 0 | na | na | 4.9167 | na | 3.5556 | 4.3000 | 4.1400 | na |
| Handan | 9.9700 | 4.7800 | 9.7700 | 4.5000 | na | 0 | 8.4800 | 5.4600 | na | 3.4000 | 4.9000 | 3.8667 | 3.3000 |
| Xingtai | 8.1200 | 4.1400 | 11.1900 | 3.5000 | na | 8.4500 | 0 | 5.3800 | 3.3000 | na | 3.2750 | 4.0000 | 3.8500 |
| Baoding | 17.0100 | 7.3700 | 13.2700 | 4.8900 | 4.1333 | 5.4800 | 5.3500 | 0 | 4.3800 | 3.1900 | 5.3800 | 6.4200 | 3.8111 |
| Zhangjiakou | 12.4700 | 4.5400 | 4.4500 | 3.3700 | 3.4000 | 3.8000 | na | 4.4300 | 0 | 2.5000 | 3.9500 | 3.3300 | 3.3000 |
| Chengde | 9.4600 | 3.7500 | 4.0100 | 4.6900 | 3.1700 | na | na | 3.3800 | 3.0000 | 0 | 3.8000 | 3.0400 | na |
| Cangzhou | 8.6100 | 8.8700 | 6.7900 | 4.2900 | 4.0333 | 4.8000 | 3.4111 | 5.5700 | 2.6000 | 2.3750 | 0 | 4.7800 | 6.0600 |
| Langfang | 19.3800 | 9.6700 | 6.1200 | 5.0400 | 4.1200 | 3.7000 | 3.7000 | 6.6400 | 3.3700 | 3.1100 | 4.9700 | 0 | 3.4125 |
| Hengshui | 7.2000 | 4.6667 | 7.8100 | 3.1889 | 3.4000 | 3.3500 | 4.1000 | 3.9400 | 3.7333 | 2.2000 | 5.7600 | 3.4625 | 0 |

**Note**: This is the railway passenger flow matrix after data cleaning. "na" represents missed data points. See the File 1 of Supplementary Files for the whole data sets.

During the study period, railway passenger flows are more than highway passenger flows in every city. It showed that railways are the main mode of intercity migration in the Beijing-Tianjin-Hebei region. In other words, the people in that region preferred railways to highways for intercity traveling. According to the *Statistical Bulletin of the development of the Chinese Transportation Industry* issued by the Ministry of Transport of the People's Republic of China, the railway passenger turnover had exceeded the highway from 2014 to 2018, and its market share was still growing (http://xxgk.mot.gov.cn/jigou/zhghs/201904/t20190412_3186720.html). The passenger turnover is the product of the number of passengers and the distance traveled. Railways had been considered as the focus of the future development of intercity passenger transportation in China long before.

Because Tencent location big data only provided top 10 inflows and top 10 outflows for every city, the passenger flow matrix is not complete. The missing values are too small to be in the top 10. Marking the missing values on the map and using arrows indicating directions, we can generate Figure 2. The arrows are like a triangle with three northern cities in the region as the base and two southern cities as the apex. It could be seen that vertical intercity connections are weaker than horizontal intercity connections, mainly because the northern and southern areas are not closely connected. That is related to the fact that the vertical span of the region is greater than the horizontal span. The Euclidean distance between the northmost Chengde city and the southmost Handan city in the region is 565 km, while the Euclidean distance between the westmost Zhangjiakou city and the eastmost Qinhuangdao city is 410 km. The weaker connections between the north and south of the Beijing-Tianjin-Hebei region are the inevitable consequence of its larger spatial span. This complies with Tobler's first law of geography, which said "everything is related to everything else, but near things are more related than distant things" (Tobler, 1970). It should be noted that though there are arrows in both Figure 2 and Figure 5, they're different. The former stands for missing values, and the latter stands for prediction outliers. The missing values are not equal to prediction outliers. The missing values are due to the limit of data sources and met with Tobler's first law of geography. Prediction bias reflects the gap between reality and expectation and is subject to random disturbances. Prediction outliers are extreme or abnormal biases. We especially care about prediction outliers. Which city pair generated outliers? And what reality questions are behind them?

    The missing values of highway passenger flows are more than railway passenger flows, that's because the spatial scales of the two transportation modes are different. Highways have a smaller spatial scale and denser tracks than railways. For example, Xilin Gol League, which is located in the middle of Inner Mongolia of China, ranked $6^{th}$ in the top 10 road passenger inflows of Chengde city on April 12 but ranked after the top 10 railway passenger inflows. Highways can connect Chengde city and Xilin Gol League directly, while railways cannot. There are more cities not belong to the Beijing-Tianjin-Hebei region such as Xilin Gol League in the top 10 highway passenger flows than railway passenger flows, resulting in more missing values of the Beijing-Tianjin-Hebei region. There are some differences in the arrows of the missing values of highway and railways but they are mainly vertical (See attached File 1).

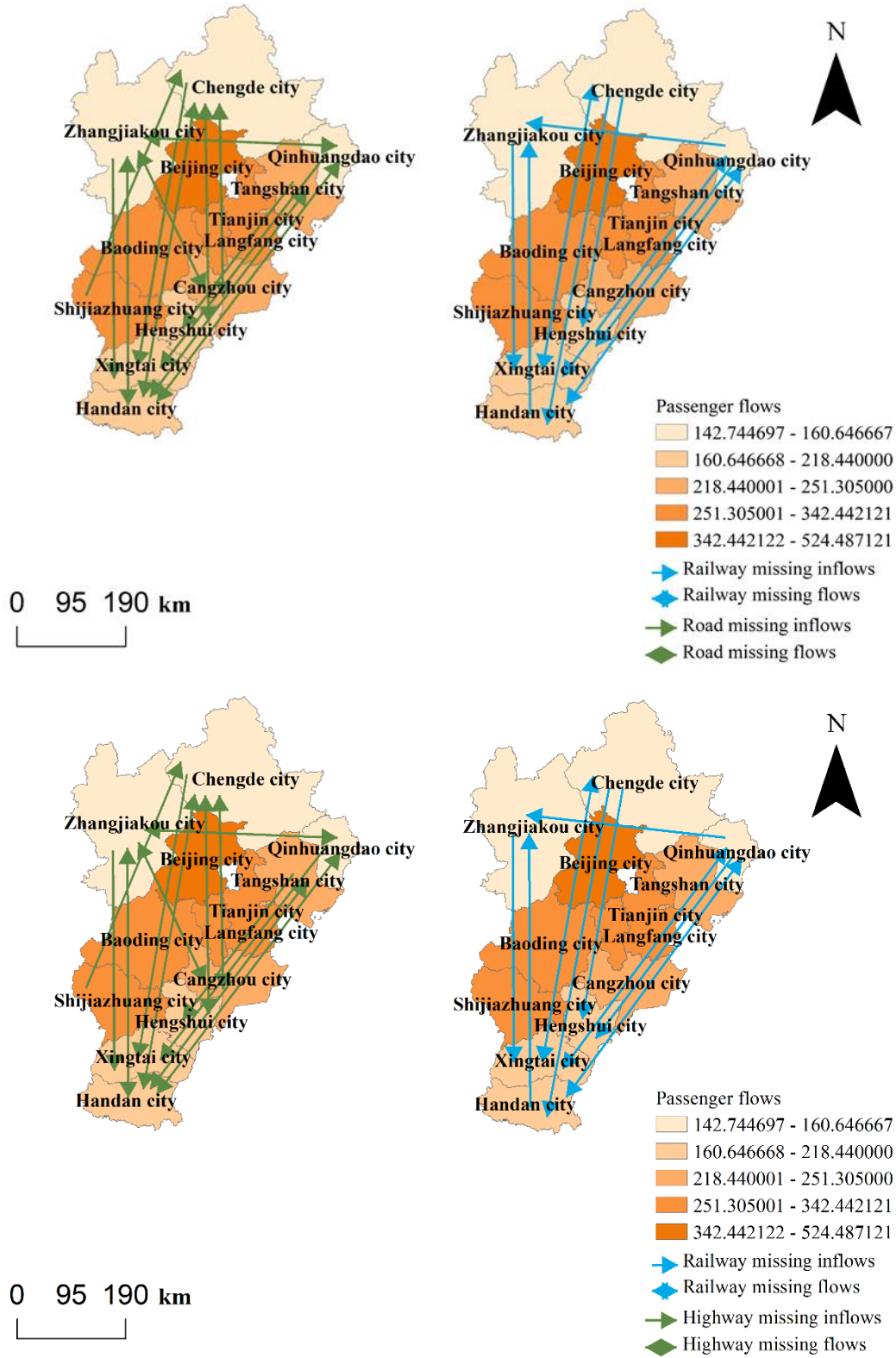

**Figure 2 Missing values in the Beijing-Tianjin-Hebei region**

**Note**: The directions of the single arrows are the same as passenger flow directions. Double arrows indicated both inflows and outflow. The little blank area beside Beijing city is Xianghe county, which is an enclave of Langfang city and its data was missing.

## 3.2 Results of the dual gravity modeling

The combination of model and data pay the foundation for our spatial analysis. On the one hand, data generate stylized facts and put constraints on models, on the other hand, models are essential to comprehend the processes at play and how the system works (Louf and Barthelemy, 2014). The model parameters can be estimated by means of least squares regression analysis based on log-linear relationships and above processed data (See File 2 for calculation processes). For railway passenger flows, the log-linear fitting results of the dual gravity model are as follows

$$\ln \hat{T}_{ij} = -2.4228 + 0.2374 \ln P_i + 0.2176 \ln P_j - 0.4412 \ln r_{ij}. \tag{7}$$

where the hat "^" indicates estimated value. It is the same below. Converting equation (7) into the general form as

$$\hat{T}_{ij} = 0.0887 P_i^{0.2374} P_j^{0.2176} r_{ij}^{-0.4412}. \tag{8}$$

The dual expression of equation (8) is as follows

$$\hat{T}_{ji} = 0.0887 P_i^{0.2176} P_j^{0.2374} r_{ji}^{-0.4412}. \tag{9}$$

Using equations (4) and (5), we can estimate the values of the gravity coefficient $G$ and distance exponent $b$, and the results are as below:

$$G = 0.0887^{2/(0.2376+0.2176)} = 0.00002373, \quad b = \frac{2*0.4412}{0.2374+0.2176} = 1.9391. \tag{10}$$

Substituting the parameter values into equation (3) yields a standardized gravity model, i.e., pure gravity model based on railway passenger flows as follows

$$\hat{I}_{ij} = 0.00002373 \frac{P_i P_j}{r_{ij}^{1.9391}}. \tag{11}$$

The value of $b$ comes between 1 and 3, indicating a fractal structure of the railway passenger flow system. For highway passenger flows, the fitting results were as follows,

$$\ln \hat{T}_{ij} = 3.3689 + 0.1390 \ln P_i + 0.1195 \ln P_j - 1.0743 \ln r_{ij}. \tag{12}$$

Transforming equation (12) into the general form as

$$\hat{T}_{ij} = 29.0477 P_i^{0.1390} P_j^{0.1195} r_{ij}^{-1.0743}. \tag{13}$$

The dual expression of equation (13) is as follows

$$\hat{T}_{ji} = 29.0477 P_i^{0.1195} P_j^{0.1390} r_{ji}^{-1.0743}. \tag{14}$$

The gravity coefficient $G$ and distance exponent $b$ can be approximately evaluated as below:

$$G = 29.0477^{2/(0.1390+0.1195)} = 210010627962.3300, \quad b = \frac{2*0.1.0743}{0.1390+0.1195} = 8.3135. \quad (15)$$

Inserting the parameter values into equation (3) yields a pure gravity model based on highway passenger flows as below:

$$\hat{I}_{ij} = 210010627962.3300 \frac{P_i P_j}{r_{ij}^{8.3135}}. \quad (16)$$

The value of $b$ departs from the reasonable range, 1 and 3, suggesting a broken fractal structure of highway passenger flows system.

As can be seen, the dual gravity models can be well fitted to the spatial flow data of interurban passengers in our study area. Both the railway passenger flows and highway passenger flows can be described with the fractal gravity models. The main test statistics of the railway model and the highway model can be summarized in Table 3. In a word, the two models can pass tests under a 99% confidence level. However, the parameter meanings are different. Comparing the two models, we can find the following characters. First, the positivity and negativity of parameters are reasonable. Population sizes have positive effects on intercity flows and distances have negative effects on intercity flows. Next, the absolute value of the coefficient of $\ln r_{ij}$ in the railway model is close to twice that of $\ln P$. While in the highway model, the ratio is close to 8 times. This shows that distance decay effects in highway transportation are more significant than that in railway transportation.

Table 3. The test statistics of the railway model and the road model

| Parameters | Railway passenger flows | | Road passenger flows | |
|---|---|---|---|---|
| **Multiple $R$** | $R$=0.8620 | | $R$=0.8967 | |
| **Standard error** | $S$=0.2560 | $C.V$=0.1483 | $S$=0.2881 | $C.V$=0.2074 |
| **$F$-statistics** | $F$=134.9560 | Sig. $F$=4.0037E-41 | $F$=173.7161 | Sig. $F$=9.1214E-45 |
| **$t$-statistics** | $t_1$=12.3668 | $P$-value=3.3208E-24 | $t_1$=6.2911 | $P$-value=4.6746E-09 |
| | $t_2$=11.2939 | $P$-value=1.9777E-21 | $t_2$=5.3680 | $P$-value=3.6536E-07 |
| | $t_3$=-10.5140 | $P$-value=2.0378E-19 | $t_3$=-21.0341 | $P$-value=3.3413E-43 |

The distance decay exponent of gravity models bears the property of a fractal parameter. The $b$ value was proved to equal the product of the Zipf exponent and fractal dimension of central place networks (Chen, 2011), that is

$$b = qD_f = D_P \to 2, \qquad (17)$$

where *q* refers to the Zipf exponent of city size distributions (Carroll, 1982; Zipf, 1949), $D_f$ denotes the fractal dimension of central place networks (see Christaller (1933) for central place theory), and $D_p$ is the average dimension of the urban population (Chen, 2015). The fractal dimension of central place networks is actually the similarity dimension of urban hierarchy (Chen and Huang, 2018a). In theory, the Zipf exponent approaches to 1, and the similarity dimension of central place networks changes around 2. Therefore, the distance exponent is about 2. The distance exponent based on the railway passenger flows is about *b*=1.9393, which is close to 2. In contrast, the distance exponent value based on the highway passenger flows is around *b*=8.3256, which departs significantly from 2. This suggests that the action at a distance of highway passenger flows is significantly weaker than that of railway passenger flows. The fractal patterns of passenger flows may be determined by their supportive system, including population distribution and transportation network. The spatial distribution of population proved to bear fractal properties (Appleby, 1996). More empirical analysis shows that the traffic network has a fractal structure, the fractal studies of road systems involve interurban traffic networks (Chen and Liu, 1999; Valério *et al*., 2016), intra-urban roads and streets (Frankhauser, 1990; Lu and Tang, 2004; Prada *et al*., 2019; Rodin and Rodina, 2000; Wang *et al*., 2017), suburban and exurban areas (Benguigui and Daoud, 1991; Sahitya and Prasad, 2020). Population patterns influence Zipf's distribution of city sizes, and traffic patterns influence networks of cities. According to equation (17), the fractal structure and parameters of spatial passenger flows depend on both the city rank-size distribution and the network of cities. The mechanism is not clear for the time being and remains to be explored deeply in future studies.

The dual gravity model can be used to predict passenger flows in our study area, so as to make up for the missing data. In fact, all mathematical modelings have two major aims: explanation and prediction (Batty, 1991; Fotheringham and O'Kelly, 1989; Kac, 1969). Using equations (8) and (9), we can predict the railway passenger flows, and using equations (13) and (14), we can predict the highway passenger flows. On the whole, the predicted values are close to the observed values (Table 4). The difference values between Table 2 and Table 4 are the prediction residuals. The residuals can be used to analyze the characteristics and changeing trends of the traffic pattern in the Beijing-Tianjin-Hebei region. What is more, the pure gravity models, equations (11) and (16), can be employed to calculate the spatial connection strengths based on railway and highway passengers flows between any two cities in the study area (See File 1).

Comparing the predicted railway passenger attractive forces based on equation (11) and the actual attractive forces, we can find the differences between theoretical expectation and reality. The calculated results were drawn in Figure 3. The actual railway passenger attractive forces network in the Beijing-Tianjin-Hebei region takes Beijing city as the first-level single core, Tianjin city and Shijiazhuang city as the secondary cores. The main axis of the actual network is the southeast and the southwest starting from Beijing city. The predicted railway passenger flow network takes Beijing city and Tianjin city as the first-level dual cores, and the entire network is distributed radially from the center, Beijing city, to the surroundings. The actual railway passenger flow values and predicted values are shown in Figure 4 at the same time. The cities whose actual flows are significantly greater than predicted values included Beijing city, Shijiazhuang city, and Baoding city. These three cities happen to form the southwest axis starting from Beijing named 'the Beijing-Baoding-Shijiazhuang development axis.' The cities whose actual flows are significantly lower than predicted values included Tianjin city, Tangshan city, and Chengde city. These three cities all lie in the northeast of the region. *The outline of the plan for Coordinated Development of the Beijing-Tianjin-Hebei region* issued in 2015 clearly stated a development framework consisting of 'one core, two main cities, three axes, four districts, and multiple nodes.' Among them, 'one core' refers to the capital, Beijing city. 'Two main cities' refer to Beijing city and Tianjin city, and Beijing-Tianjin linkage is expected to be the development engine of the region. 'Three axes' refer to Beijing-Tianjin axis, Beijing-Baoding-Shijiazhuang axis, and Beijing-Tangshan-Qinhuangdao axis. These three axes will serve as industrial regions and metropolitan areas. It could be seen from Figure 3(c) that the 'one core', Beijing city, is more important in the region than predicted. Tianjin city as one of the 'two main cities' has a close connection with Beijing, but its connections with other cities are weaker than predicted, especially the connection with Tangshan city. Among the 'three axes', the actual connection strength of the Beijing-Tianjin axis is close to the predicted value. The actual connection intensity of the Beijing-Baoding-Shijiazhuang axis is significantly stronger than predicted. The actual connection intensity of the Beijing-Tangshan-Qinhuangdao axis is lower than predicted, and the attractive force between Beijing city and Tangshan city is significantly lower than predicted.

**Table 4. Partial predicted results for railway passenger flows of Beijing-Tianjin-Hebei region (average values)**

| Predicted flows | Beijing | Tianjing | Shijiazhuang | Tangshan | Qinhuangdao | Handan | Xingtai | Baoding | Zhangjiakou | Chengde | Cangzhou | Langfang | Hengshui |
|---|---|---|---|---|---|---|---|---|---|---|---|---|
| Beijing | 0 | 19.1492 | 10.2015 | 13.2289 | 8.0987 | 7.3460 | 6.5382 | 11.0226 | 10.4151 | 8.8789 | 8.3132 | 16.2535 | 7.2345 |
| Tianjing | 18.9494 | 0 | 9.1311 | 13.7065 | 7.7131 | 6.8659 | 6.0687 | 9.4598 | 7.1268 | 6.9857 | 10.4808 | 12.1607 | 7.1573 |
| Shijiazhuang | 9.8257 | 8.8874 | 0 | 5.8317 | 4.0055 | 7.0753 | 7.1020 | 7.5049 | 4.9576 | 3.7760 | 5.0901 | 5.1740 | 6.7488 |
| Tangshan | 12.7711 | 13.3717 | 5.8453 | 0 | 7.4752 | 4.5374 | 3.9666 | 5.6602 | 5.0783 | 6.2679 | 5.4438 | 7.1379 | 4.3785 |
| Qinhuangdao | 7.6452 | 7.3581 | 3.9259 | 7.3096 | 0 | 3.1499 | 2.7243 | 3.6137 | 3.4189 | 4.4213 | 3.3853 | 4.0426 | 2.9003 |
| Handan | 6.9814 | 6.5941 | 6.9814 | 4.4668 | 3.1711 | 0 | 8.4986 | 4.5778 | 3.5164 | 2.8762 | 3.8086 | 3.6970 | 4.8192 |
| Xingtai | 6.1198 | 5.7402 | 6.9018 | 3.8459 | 2.7012 | 8.3701 | 0 | 4.1478 | 3.0796 | 2.4757 | 3.3283 | 3.2361 | 4.4406 |
| Baoding | 10.4262 | 9.0423 | 7.3704 | 5.5459 | 3.6209 | 4.5562 | 4.1916 | 0 | 4.6461 | 3.5230 | 4.9283 | 5.6018 | 5.0238 |
| Zhangjiakou | 9.8378 | 6.8027 | 4.8619 | 4.9688 | 3.4209 | 3.4949 | 3.1077 | 4.6396 | 0 | 3.7653 | 3.3052 | 4.4719 | 3.1907 |
| Chengde | 8.3017 | 6.6004 | 3.6655 | 6.0704 | 4.3790 | 2.8296 | 2.4730 | 3.4824 | 3.7271 | 0 | 2.9744 | 4.0851 | 2.6173 |
| Cangzhou | 7.7470 | 9.8701 | 4.9249 | 5.2549 | 3.3419 | 3.7346 | 3.3137 | 4.8554 | 3.2608 | 2.9646 | 0 | 4.5373 | 4.3148 |
| Langfang | 15.2916 | 11.5617 | 5.0540 | 6.9562 | 4.0289 | 3.6598 | 3.2528 | 5.5717 | 4.4542 | 4.1106 | 4.5808 | 0 | 3.7157 |
| Hengshui | 6.7381 | 6.7365 | 6.5262 | 4.2243 | 2.8615 | 4.7229 | 4.4187 | 4.9468 | 3.1462 | 2.6072 | 4.3125 | 3.6784 | 0 |

**Note**: This is the predicted railway passenger flow matrix based on data cleaning results. The missing values are made up. See File 1 of Supplementary Files for the whole predicted values.

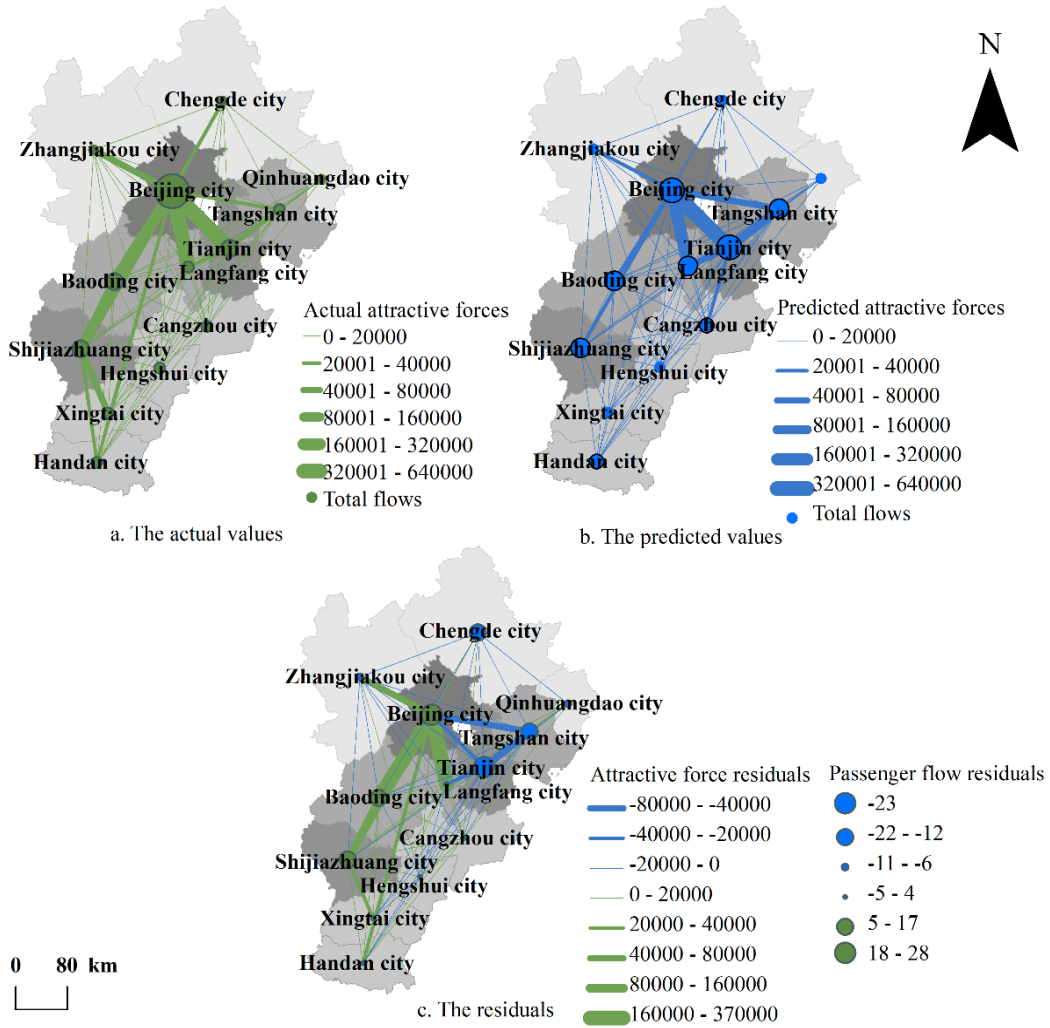

**Figure 3 The actual values, predicted values, and residuals of regional railway passenger forces**

**Note:** The thickness of straight lines in this picture reflects the magnitude of attractive forces between corresponding cities. Actual attractive forces in Fig. a are calculated based on actual flows $T_{ij}$ and $T_{ji}$ by substituting the predicted values of parameters $u$ and $v$ in equation (8) into equation (3). Predicted attractive forces in Fig. 3(b) are directly calculated based on equation (11). In other words, actual attractive forces in Fig. 3(a) are calculated based on actual

flows, while predicted attractive forces in Fig. 3(b) are calculated based on predicted flows. Residuals in Fig. 3(c) are differences between actual values and predicted values. The distribution of attractive forces and residuals reflects the spatial pattern of regional transportation.

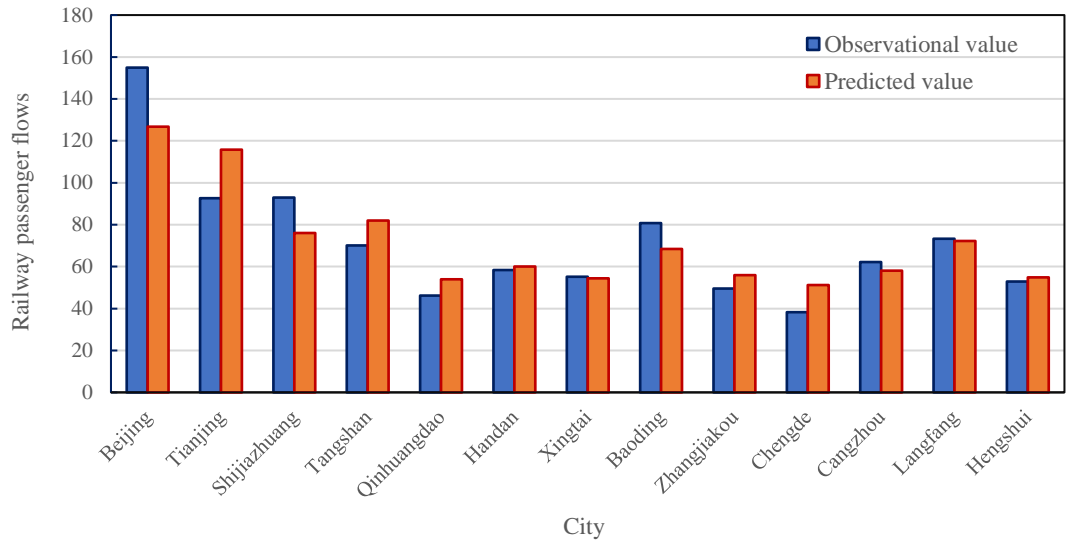

**Figure 4 Comparison of actual values and predicted values of railway passenger flows**

The global consistency and local differences between the prediction values based on gravity models and the observed values of passenger flows can reflect changing trends of spatial patterns in the urban system. In terms of the Beijing-Tianjin-Hebei region, the past spatial pattern with the Beijing-Tianjin-Tangshan triangle as the center is changing to a new pattern with the Beijing-Shijiazhuang-Tangshan big triangle as the center. That big triangle concludes Tianjin city and Baoding city. From the perspective of observed values and predicted values, the passenger flows in Beijing city, Tianjin city, and Tangshan city remain absolutely prominent throughout the study area. However, if changing trends are considered, passenger flow connections between Tianjin city, Tangshan city, and other cities show signs of relatively decreasing, while passenger flows between Beijing city, Shijiazhuang city, Baoding city, and other cities are increasing. The prediction residuals of Tianjin city and Tangshan city with other cities are mostly negative. The sum of prediction residuals for these two cities is significantly negative. In contrast, the prediction residuals of passenger inflows and outflows in Beijing city, Shijiazhuang city, and Baoding city are mostly positive. The sum of prediction residuals for these three cities is significantly positive. It means that the external connections of Tianjin city and Tangshan city are lower than expected while the external connections of Beijing city, Shijiazhuang city, and Baoding city are higher than expected. Though the population size of Beijing city is smaller than Shanghai city, Beijing city is the primate city in the sense of politics and function. The passenger flow connections of the Beijing-Baoding-Shijiazhuang axis, which runs northeast-southwest, are gradually strengthening. This may be influenced by China's policy of developing the Beijing-Tianjin-Hebei region and is also related to the impact of Beijing-Guangzhou Railway. Tianjin city is the gateway for Beijing city to open up to the outside and the development pattern of the Beijing-Tianjin-Tangshan triangle is a product of China's Reform and Opening up. The overall pattern in the region seems to be changing from outward-oriented development to inward-oriented development.

## 4 Discussion

The results of data analysis shows that dual gravity models are available to describe the spatial pattern of railway and highway passenger flows in the Beijing-Tianjin-Hebei region. The goodness of models fitted to observational data is satisfying. The predicted values of models roughly match the actual values. If the gravity models based on flows are transferred to the standard gravity model, it is found that the railway passenger flow model is different from the highway passenger flow model significantly where parameter values are concerned. The standardized distance exponent of the railway gravity model is close to 2, which is not far from the predicted value in theory (Chen, 2015; Chen and Huang, 2018a). However, the standardized distance exponent of the highway gravity model is more than 8, which is more than 4 times the railway distance exponent. That means, the railway passenger flows in the Beijing-Tianjin-Hebei area decay normally with normal fractal structure, while the highway passenger flows quickly decay and deviate from the normal fractal structure. Fractal suggests a kind of optimized structure in nature (Lin and Li, 1992). A fractal object can occupy its space in the best way. The departure of an urban phenomenon or a traffic network from a fractal structure suggests a kind of evolution fluctuation or development obstacle. The railway passenger flows show the long-distance effect and the highway passenger flows show the localization property to some extent. People choose highway transport only for short trips. For long trips, they choose railway transport mostly. It can be seen that railway passenger flows follow the first law in geography better than the highway passenger flows. A series of spatial analyses can be conducted by using the models. First, the models can be used to predict the passenger flows and complement the small part of missing data approximately. Second, the fractal structure and abnormal fractal structure of the spatial flows can be revealed. Third, the differences between predicted values and actual values of passenger flows can be compared using the models. The future evolutionary trends of transport patterns in the Beijing-Tianjin-Hebei region can be judged by the residuals of prediction.

The dynamical mechanism behind the outliers should be discussed. In Section 3.2, the spatial pattern of the prediction bias was depicted. The residuals that are not within plus or minus two standard deviations will be discussed next. There is only one outlier in the highway passenger flow model. The standard deviation of highway passenger flow from Beijing city to Baoding city is 2.04. The railway passenger flow model has more outliers than the highway model (Figure 5(c)). The spatial pattern of railway outliers is highly related to the unbalanced distribution of the regional intercity railway network. Outliers that are beyond two standard deviations appear in the middle of the region, where there are densely distributed railways. Outliers that are below two standard deviations appear in the north of the region, where there are mountains and sparsely distributed railways. In contrast, the distribution of the regional highway network is more even, resulting in fewer outliers in the highway model.

Next, the accessibility concept and intervening opportunities will be considered. Accessibility can be used as an indicator that reflects the potentials of opportunities for interaction (Hansen, 1959). Accessibility can be calculated based on node adjacency, the mass of cities, distances, travel costs, travel time, utility indexes, and the number of passengers (Bruinsma and Rietveld, 1998). The fastest travel time from node *i* and node *j* can be seen as the accessibility between those nodes, which is $A_{ij}$. (Cao and Yan, 2003). Based on the data of the 12306 website (https://www.12306.cn), which is China's most authoritative railway ticketing website, the accessibility matrix of the Beijing-Tianjin-Hebei region was collected. The relationships between railway model residuals and population, distance, and node accessibility are drawn as scatter plots (Figure 6). As can be seen, there was is no obvious trend in the scatter plots of population and distance. That means there were are no structural problems in the model. Although there is not an absolute quantitative relationship between node accessibility and residuals, accessibility will affect the upper and lower bounds of residual fluctuations. The better the accessibility, the more the upper bound of residual fluctuation tends to exceed the plus two standard deviations. The worse the accessibility, the more likely the lower bound tends to exceed the minus two standard deviations. After introducing the accessibility into the railway model for log-linear fitting, all negative outliers disappear. It can be inferred that negative outliers are related to differences in node accessibility.

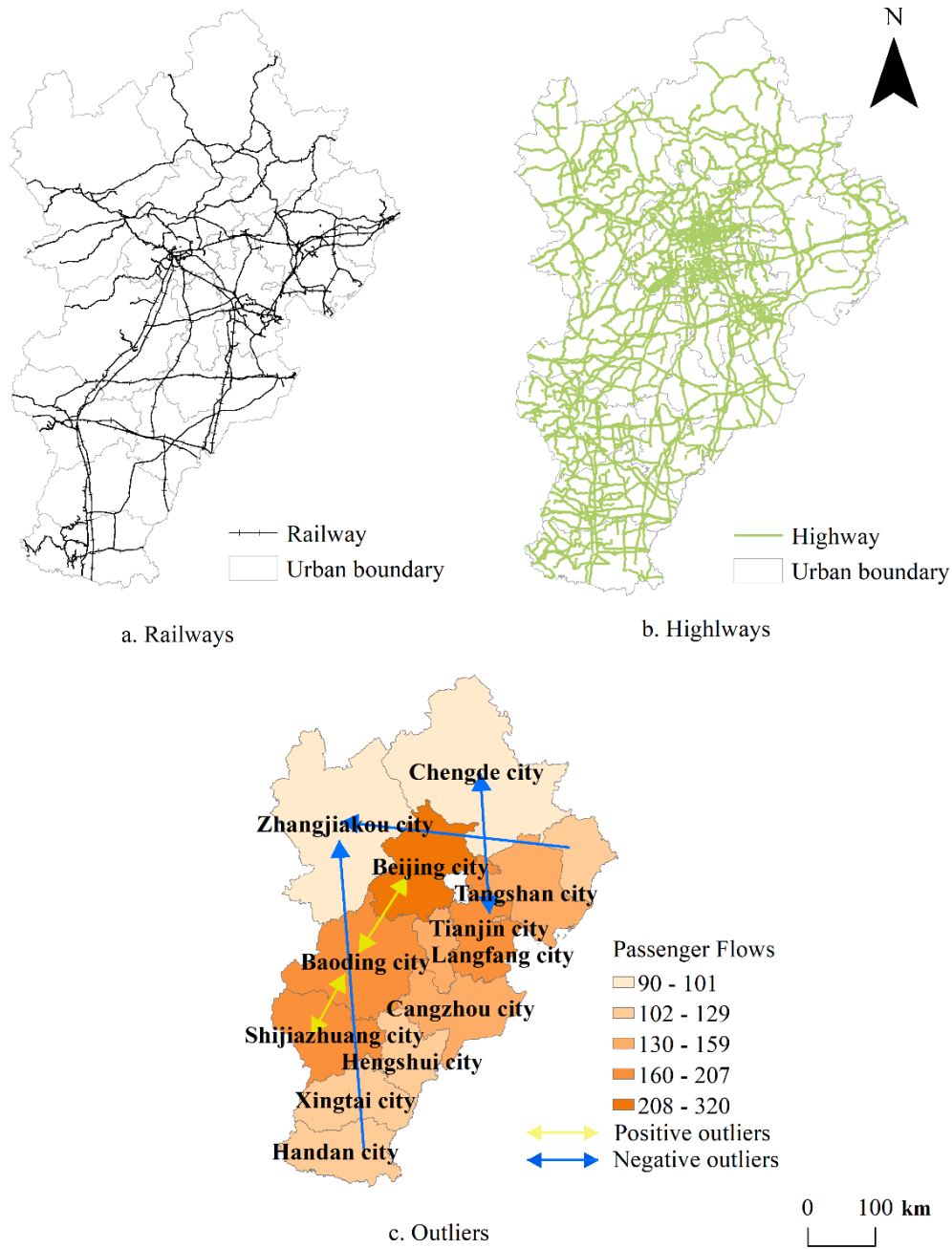

**Figure 5 The railways, highways, and railway passenger flow outliers in the Beijing-Tianjin-Hebei region**

The occurrence of positive outliers is related to the differences in intervening opportunities. S. A. Stouffer believes that distance does not constitute the substantial cause of flows' decay in space, but the real reason lies in the increase of intervening opportunities (Stouffer, 1940). The imbalance of the railway network will magnify the impact of intervening opportunities. The number of cities passed by the railway in the shortest travel time between origins and destinations is recorded as intervening opportunities. After introducing intervening opportunities in the railway model, positive outliers obviously decrease.

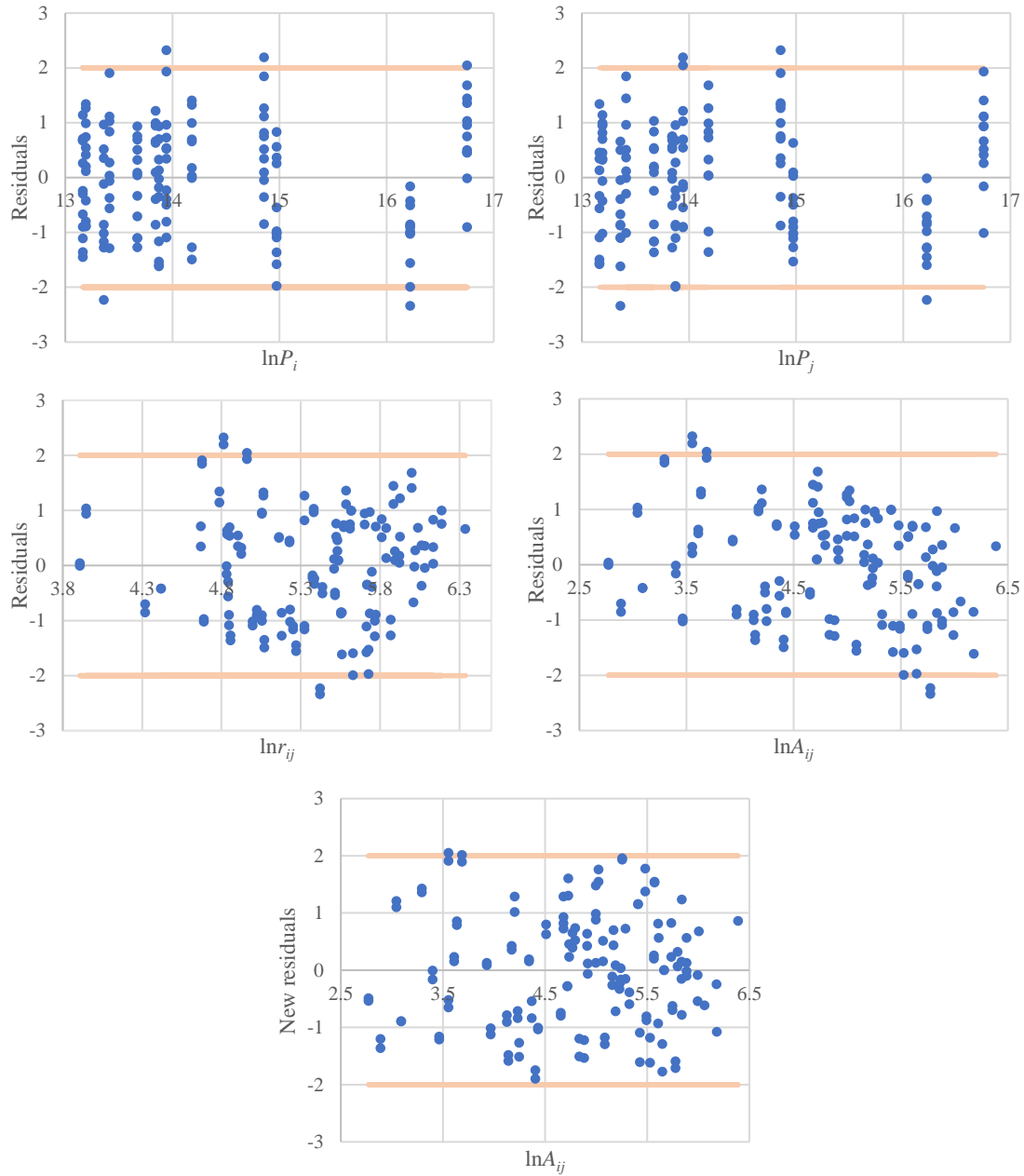

**Figure 6 The scatter plots before and after introducing accessibility in the model**
**Note:** New residuals are calculated after introducing $A_{ij}$ to the railway model. The calculation method is a least squares regression analysis based on log-linear relationships between $T_{ij}$, $P_i$, $P_j$, $r_{ij}$, and $A_{ij}$. Residuals have been standardized.

The flows between cities are related to regional gravity and spatial interaction. The common models for predicting population flows are Wilson's (1970) spatial interaction models. However,

Wilson's models are not applicable to this study. First, one of the prerequisites of Wilson's models is data integrity. It's desirable to have complete population flows between any two cities. Taking a step back, we have to at least get the total inflows and outflows of every city to build the models. However, we mined most of the flow data between cities and obtained neither exact total inflows nor outflows of every city nor total population flows of all cities. Therefore, Wilson's models (total-flow constrained model, singly constrained model, doubly constrained model) couldn't be used sufficiently in our study. Second, Wilson's models are essentially nonlinear programming models to optimize traffic flows. The models are very useful for urban planning and regional transportation planning. In a word, Wilson's spatial interaction models apply to normative research aimed at urban and transportation planning and the gravity model applies to behavioral research oriented to empirical analysis. When transportation flow data is incomplete, dual gravity models can predict flows in urban space indirectly using urban attractive forces and avoid total population flow data of every city.

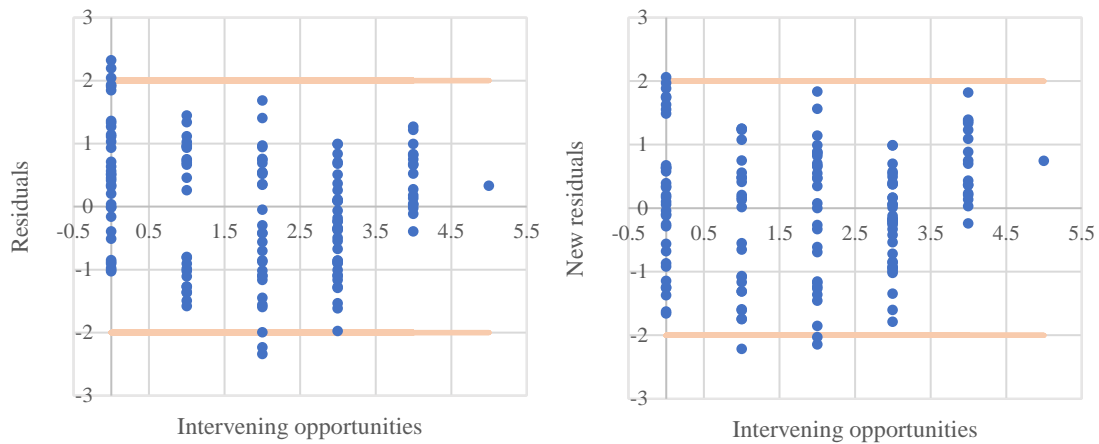

**Figure 7 The scatter plots before and after introducing intervening opportunities in the model**
**Note:** The calculation method in Figure 7 is the same as Figure 6. When we perform the regression, all intervening opportunities need to be added 1 before being logged. Because the intervening opportunity for neighboring cities is just 0, and ln0 cannot be calculated.

The novelty of our study lies in three aspects. First, in the case of missing a very small part of spatial data, that is, the spatial dataset is not complete, make up the grid data and reveal the spatial distribution characteristics of passenger flows in the study area using dual gravity models. Second, the ideas from fractal theory were employed to reveal the deep structure of the traffic network. Third, the prediction residuals were used to analyze the evolutionary trends of the spatial pattern of urban systems in the study area. The shortcomings of the study are in two areas. First, the urban population size data and population flow data for analysis were not an exact match temporally. This defect had no significant effects on population flow analysis because of the following two aspects. On one hand, the urban population size is relatively stable. Our research relied heavily on the relative sizes between cities instead of absolute sizes. The relative sizes between cities in the Beijing-Tianjin-Hebei region will not change much in a few years. On the other hand, urban sizes have a kind of time-lag effect on population flows. If the interaction between urban sizes is considered as inputs and the population flows are considered as outputs, there is a time lag between sizes (inputs) and flows (outputs). That is known as the response delayed effect reflected on spatial interaction in geographic space (Chen, 2009). Second, the local gravity analysis was not carried out. The dual gravity models used in our study are global models to describe urban attractive forces based on traffic data. The models can characterize the traffic flow pattern in a region from a holistic perspective. However, if researchers want to reveal the spatial-temporal evolutionary characteristics of population flows in the study area systematically and thoroughly, it's necessary to use local gravity models to explore the spatial heterogeneity of urban population migration behavior. What's more, the dual gravity models are a kind of fractal gravity models in essence. The fractal characteristics of population migration in the Beijing-Tianjin-Hebei region require further explorations systematically.

## 5 Conclusions

The dual gravity model fits the railway and highway passenger flows well with a few prediction outliers in the Beijing-Tianjin-Hebei region. The highway model has almost no outliers, and the distribution of highways in the study area is relatively even. The prediction outliers of spatial pattern of railway model is highly related to the distribution of the regional railway network in the Beijing-Tianjin-Hebei region. The negative outliers come mainly from the differences in node accessibility. The positive outliers come mainly from the differences in intervening opportunities in the urban system. Differences in node accessibility and intervening opportunities are both related to the spatial heterogeneity of the regional railway network. If node accessibility and intervening opportunities are introduced to the model separately, outliers will be effectively eliminated. However, the research goal of this paper is not to improve the mathematical models according to the exceptional phenomena, but to reveal the real problems by using dual gravity models. Based on the above results and discussion, the main conclusions could be drawn as follows.

**First, dual gravity models can effectively describe and predict the spatial pattern of passenger flows between cities.** In terms of railway and highway passenger flows in the Beijing-Tianjin-Hebei region, observed values and predicted values are overall consistent with each other, and some local differences in them can reflect gaps between reality and expectation. Partial missing data can be made up by predicted values. Then spatial interaction analysis based on entropy maximization can be conducted using this complete dataset to provide planning guides for future transportation development in the study area. There is a key problem that can be illustrated by gravity modeling effects of the regional passenger flow distribution in the study area: the distance-decay effect still dominates the changing trends of spatial patterns in human geographic systems. Due to rapid changes in new technologies, many scholars doubt whether the geographic distance-decay law is still valid. The modeling effects in our research gave a positive answer to the question. The distance exponent in the railway model is very close to the theoretical expectation, which indicates that the roles of railways in the transportation system of China mainland have characteristics of large-scale and stability. In contrast, the distance exponent in the highway model is more than 4 times the theoretical expectation, which indicates short-range effect and the poorer stability of highway passenger flows.

**Second, dual gravity models reflect a fractal pattern and variation characteristics of passenger flow distribution in the study area.** Dual gravity models are essentially fractal models and normalized distance exponents are fractal parameters. These parameters reflect the spatial pattern of urban population distribution, linking both Zipf's rank-size distribution and central place networks. Mathematical models mirror macro patterns while model parameters mirror behaviors of micro elements. In terms of the Beijing-Tianjin-Hebei region, the railway passenger flows satisfy the gravity model based on inverse power law. The normalized distance exponent value of the railway model comes between 1 and 2, representing a fractal dimension. It can be seen that railway passenger flows in the region are consistent with fractal patterns at both macro and micro levels. The highway passenger flows also satisfy the gravity model based on inverse power law, which shows a macrostructure consistent with fractal patterns. However, the parameter reflecting micro-level element interactions in the highway model deviates from a reasonable range. As a fractal

parameter, the distance exponent in the highway model should be between 0 and 3. In reality, that distance exponent is greater than 8, which is ridiculously high. The high distance exponent means a high distance-decay effect apparently. However, it exceeds the reasonable theoretical range and shows a deviation of railway passenger flows from a fractal structure for the time being. The reason for that phenomenon needs to be further explored.

**Acknowledgements**

This research was sponsored by the National Natural Science Foundation of China (Grant No. 42171192). The supports are gratefully acknowledged.

## Supplementary Files

**[Supplementary File 1]** *Spatial data sets for building dual gravity models of interurban passenger flows (Excel)*. This file includes the data of railway passenger flows and highway passenger flows mined from Tencent big data as well as city sizes and interurban distance. The predicted values of spatial flows by the dual gravity models and related residuals are also shown in this file.

**[Supplementary File 2]** *Spatial modeling processes by fitting dual gravity models to processed data (Excel)*. This file displays the processes and results of multivariate regression modeling for both the railway passenger flows and highway passenger flows. Readers can verify our results by repeating our calculations, or imitating these operations to carry out their own research.

Press

Batty M, Morphet R, Masucci, Stanilov K (2014). Entropy, complexity, and spatial information. *Journal of Geographical Systems*, 16: 363-385

Benguigui L, Daoud M (1991). Is the suburban railway system a fractal? *Geographical Analysis*, 23: 362-368

Bruinsma F, Rietveld P (1998). The accessibility of European cities: theoretical framework and comparison of approaches. *Environment and Planning A*, 30(3): 499-521

Bussiere R, Snickers F (1970). Derivation of the negative exponential model by an entropy maximizing method. *Environment and Planning A*, 2(3): 295-301

Cao XS, Yan XP (2003). The impact of the evolution of land network on spatial structure of accessibility in the developed areas: the case of Dongguan city in Guangdong province. *Geographical Research,* 22(3): 305-312 (in Chinese)

Carroll C(1982). National city-size distributions: What do we know after 67 years of research? *Progress in Human Geography*, 6(1): 1-43

Chen YG (2009). Urban gravity model based on cross-correlation function and Fourier analyses of spatio-temporal process. *Chaos, Solitons & Fractals*, 41(2): 603-614

Chen YG (2012). The rank-size scaling law and entropy-maximizing principle. *Physica A*, 391(3): 767-778

Chen YG (2015). The distance-decay function of geographical gravity model: Power law or exponential law?. *Chaos, Solitons & Fractals*, 77: 174-189

Chen YG, Huang LS (2018a). A scaling approach to evaluating the distance exponent of the urban gravity model. *Chaos, Solitons & Fractals*, 109: 303-313

Chen YG, Huang LS (2018b). Spatial Measures of Urban Systems: from Entropy to Fractal Dimension. *Entropy*, 20(12): 2099-4300

Chen YG, Liu JS (1999). The DBM features of transport network of a district: A study on the Laplacian fractals of networks of communication lines. *Scientia Geographica Sinica*, 19(2): 114-118 (In Chinese)

Christaller W (1933). *Central Places in Southern Germany*. Englewood Cliffs, New Jersey: Prentice Hall (translated by C. W. Baskin in 1966)

Converse PD (1949). New laws of retail gravitation. *Journal of marketing*, 14(3): 379-384